\documentstyle[twoside,fleqn,espcrc2,epsf]{article}
\epsfverbosetrue

\newcommand{\AmS}{{\protect\the\textfont2
  A\kern-.1667em\lower.5ex\hbox{M}\kern-.125emS}}

\def\dfrac#1#2{{\displaystyle \frac{#1}{#2}}}

\def\veg#1{{\bf #1}}
\def\vek#1{\mbox{\protect\boldmath $#1$}}
\def\slsh#1{\rlap{$\,/$}#1}

\newcommand{\cD}{{\cal D}}

\newcommand{\half}{\mbox{\small $\frac{1}{2}$}}

\def\lsim{\;\raisebox{-.4ex}{\rlap{$\sim$}} \raisebox{.4ex}{$<$}\;}

\newcommand{\vdot}{\!\cdot\!}

\title{The Charm Quark on the Lattice%
\thanks{Talk given at
{\em Lattice '92}, Amsterdam, Sept.\ 15--19, 1992.}
\hfill {\normalsize FERMILAB-CONF-92/329-T}
}

\author{Andreas S. Kronfeld \\
\medskip
Theoretical Physics Group, 
Fermi National Accelerator Laboratory, 
Batavia, Illinois, USA}
        
\begin{document}

\begin{abstract}
We formulate lattice fermions in a way that encompasses Wilson fermions
as well as the static and non-relativistic approximations.
In particular, we treat $m_qa$ systematically ($m_q$ is the fermion
mass) showing how to understand the Wilson action as an effective action
for systems with $\vek{p}\ll m_q$.
The results show how to extract matrix elements and the spectrum from
simulations with $m_qa\approx1$, which is relevant for the charm quark.
\end{abstract}

\maketitle

\section{INTRODUCTION}
Current numerical calculations of lattice QCD have lattice spacings such
that $m_qa\not\ll 1$ for charm and bottom quarks.
To extract meaningful results one must understand how to formulate
fermions under these circumstances.
One approach is to formulate this problem starting from the large mass
limit.
Our approach is to deduce an action that contains the Wilson
\cite{Wil77}, static \cite{Eic87}, and non-relativistic \cite{Lep87}
formulations as special cases.
An important question is how much of the flexibility of our
formulation should be exploited to study, for example, quarkonia and
heavy-light systems.
A related and practical issue is how one can interpret numerical
results obtained with the Wilson action when $m_qa\approx1$.
This paper offers a brief discussion of these issues.

The key idea is that one should analyze every heavy-quark theory,
including the Wilson theory, as an effective (cutoff) theory.
After all, every numerical simulation has $a\neq0$.
In other words, one should avoid being constrained by an implicit desire
to take the continuum limit by brute force.
Naturally, the goal remains the same, to extract universal
(i.e.\ cutoff-free) information, but one would like to do so as
efficiently as possible.

With any effective theory, one must understand its limitations.
It is worth recalling that the characteristic momenta of the systems
that we would like to study are much smaller than $m_q$.
In quarkonia the typical momentum scales are $\alpha_S(m_q)m_q$ and
$\alpha_S(m_q)^2m_q$, so $pa\lsim1$ for a wide range of $m_qa$.
In heavy-light systems the typical momentum is roughly
$\Lambda_{\rm QCD}$, independent of $m_q$.
Below we argue that the Wilson action describes heavy
quarkonia with the same validity as the non-relativistic effective
theory and describes heavy-light systems with the same validity as
the non-relativistic and static effective theories.


\section{THE NEW ACTION}\label{new}
The familiar heavy-quark effective theories treat time and space
asymmetrically.
Therefore, it should not be too surprising that our result \cite{Kro92}
boils down to introducing a second hopping parameter to the familiar
Wilson action.
\begin{equation}\label{kappa-kappa}
\begin{array}{l}
S={\displaystyle \sum_x} \bar{\psi}_n\psi_n
- \kappa_0 {\displaystyle \sum_{n}} \left[\rule{0.0em}{0.97em}
\bar{\psi}_n(1-\gamma_0)U_{n,0}\psi_{n+\hat{0}}
\right. \\[1.0em] \hspace{8.5em} + \left.\rule{0.0em}{0.97em}
\bar{\psi}_{n+\hat{0}}(1+\gamma_0)U_{n,0}^\dagger\psi_n
\right] \\[1.0em] \hspace{6.0em}
-\,\kappa_S {\displaystyle \sum_{n,i}} \left[\rule{0.0em}{0.97em}
\bar{\psi}_n(1-\gamma_i)U_{n,i}\psi_{n+\hat{\imath}}
\right. \\[1.0em] \hspace{8.5em} + \left.\rule{0.0em}{0.97em}
\bar{\psi}_{n+\hat{\imath}}(1+\gamma_i)U_{n,i}^\dagger\psi_n
\right].
\end{array}
\end{equation}
This action describes massless quarks when
$\kappa_0=\kappa_S\rightarrow1/8$.
It also describes static quarks when for $\kappa_S=0$.
Intuitively, there ought to be a trajectory in the
$\kappa_0$--$\kappa_S$ plane that describes somewhat massive quarks.

Eq.~(\ref{kappa-kappa}) actually describes the simplest in a class of
actions, in which lattice artifacts are corrected by the Symanzik
improvement program.
Further details appear elsewhere \cite{Kro92}, but setting
$\kappa_0=\kappa_S=\kappa$, the most important term is the
``clover'' term \cite{She85}
\begin{equation}\label{Wilson-action}
\Delta S =
{\displaystyle \frac{ic\kappa}{2} \sum_{n,\mu,\nu}}
\bar{\psi}_n\sigma_{\mu\nu} a^2C_{n,\mu\nu}\psi_n ,
\end{equation}
where $C_{\mu\nu}$ is a lattice approximant to $F_{\mu\nu}$.
This is the form usually used in numerical work, but for perturbative
calculations and contact with continuum theories it is convenient to
write $n=x/a$, $(2\kappa)^{-1}=4+am_0$, and
$(2\kappa)^{1/2}\psi_n=a^{3/2}\psi(na)$.

\section{THE CONTINUUM}\label{continuum}
This section is in Minkowski space; everything else is in Euclidean
space.
The continuum action is
$ 
S = \int d^4x\,\bar{\psi}(i\slsh{D} + m)\psi.
$ 
Choose the gauge $A_0=0$, split the fermion into two-component fields,
and redefine the fields $\phi\mapsto e^{im_0t}\phi$
and $\chi\mapsto e^{-im_0t}\chi$.
Then, using the appendix,
\begin{equation}
\begin{array}{rr}
S = {\displaystyle \int}d^4x\,\left(
  \phi^\dagger       i\partial_0      \phi
+ \chi^\dagger   (i\partial_0 + 2m)   \chi
	\right. \\[1.0em] \left.
+ \phi^\dagger\vek{D}\vdot\vek{\sigma}\chi^\dagger
- \chi\vek{D}\vdot\vek{\sigma}\phi \right) .
\end{array}
\end{equation}
Because $\chi$ appears quadratically in the functional integral, it can
be integrated out, leaving an effective theory for $\phi$ with action
\begin{equation}
S = \int d^4x\,\phi^\dagger\left( i\partial_0-
\vek{D}\vdot\vek{\sigma}
\frac{1}{2m+i\partial_0}\vek{D}\vdot\vek{\sigma}
\right)\phi
\end{equation}
For systems for which
$i\partial_0\vek{D}\vdot\vek{\sigma}\phi\ll
m\vek{D}\vdot\vek{\sigma}\phi$
the non-local expression can be expanded
\begin{equation}
\frac{1}{2m+i\partial_0}\approx
\frac{1}{2m} - \frac{i\partial_0}{(2m)^2} -
\frac{(i\partial_0)^2}{(2m)^3} - \cdots.
\end{equation}
Higher time derivatives can be eliminated using the equations of
motion (e.g.\ $[\partial_0,\vek{D}]=\vek{E}$) \cite{Hor92},
$(\vek{D}\vdot\vek{\sigma})^2=\vek{D}^2+i\vek{\sigma}\vdot\vek{B}$,
etc.

Usually only the $\vek{D}^2/(2m)$ and the
$\vek{\sigma}\vdot\vek{B}/(2m)$ terms are treated explicitly,
with short-distance effects of the higher terms compensated by
adjusting the couplings of these two terms.
Overall, this type of analysis is very successful, for example,
motivating static potential models and developing of heavy-quark
symmetry.
One limitation is obvious: the field $\phi$ now creates states with
energy-momentum relation $E=p^2/(2m)$ instead of $E=m+p^2/(2m)$.
Hence the splitting between sectors with a different number of quarks
does not come out right.
Usually they can be corrected by hand, but one should keep an open mind
to more subtle dynamical effects that might alter, say, mixing between
heavy-quark mesons and glueballs.

\section{THE LATTICE}\label{lattice}
Suppose that a lattice theory has the following $1/M$ expansion for
the Hamiltonian:
\begin{equation}\label{M-expansion}
\begin{array}{l}
\hat{H} =
\hat{a}^\dagger\left[ M_1 - \dfrac{\vek{D}^2}{2M_2} -
\dfrac{i\vek{\sigma}\vdot\vek{B}}{2M_3} \right. \\[1.0em]
\hspace{8.0em} \left. -\dfrac{(\vek{D}^2)^2}{4M_4^3}-\cdots \right]\hat{a}
 \\[1.0em] \hspace{1.7em} + \mbox{ anti-quark terms},
\end{array}
\end{equation}
but the $M$'s are not all the same.
(Naturally, in a lattice theory the Hamiltonian is defined through the
transfer matrix formalism.)
They play the following roles:
\begin{itemize}
\item
$M_1$ counts quarks and anti-quarks \hfill *
\item
$M_2$ fixes spin-averaged splittings \hfill *****
\item
$M_3$ fixes hyperfine structure \hfill ***
\item
$M_4$ fixes fine structure \hfill **
\end{itemize}
Of these ``masses'' $M_2$ is the most important physically.
As in sect.~\ref{continuum} (where $M_1=0$ and $M_2=m$) the fact that
$M_1\neq M_2$ can be corrected by hand.
If that strategy is impractical or distasteful, one simply switches
back to eq.~(\ref{kappa-kappa}); sect.~\ref{asymmetric-section}
shows how the $\kappa$'s can be tuned so that $M_1=M_2$.
Similarly, the other masses multiply operators that are easy to
interpret physically, so their deviations could be taken into account as
perturbations.
Alternatively, they can be corrected as part of the Symanzik improvement
of eq.~(\ref{kappa-kappa}).

All actions in the class exemplified by eq.~(\ref{kappa-kappa}) take the
form of eq.~(\ref{M-expansion}), in the limit $|\vek{p}|\ll M_2$,
$pa\ll 1$, even without restriction on $aM_2$ \cite{Kro92}.
For example, at tree level the Wilson action  has
\begin{equation} \label{M1}
M_1=a^{-1}\log(1+m_0a),
\end{equation}
\begin{equation} \label{M2}
\begin{array}{l}
\hspace{-0.5em} (2M_2)^{-1}=
\dfrac{2+4m_0a+m_0^2a^2}{2m_0(1+m_0a)(2+m_0a)}\\[1.2em]
\hspace{3.6em} \approx(2m_0)^{-1},
\end{array}
\end{equation}
and
\begin{equation} \label{M3}
(2M_3)^{-1} = (2M_2)^{-1} - \frac{1-c}{2m_0(1+m_0a)}.
\end{equation}
These expressions are corrected at ${\rm O}(g_0^2)$.
The first two are easily extracted from the (free) energy-momentum
relation
\begin{equation}
\cosh(Ea) = 1 + \frac{1}{2}
\frac{(m_0a+\half a^2\hat{\vek{p}}^2)^2+a^2\vek{S}^2}%
{1+m_0a+\half a^2\hat{\vek{p}}^2} .
\end{equation}
Note that $(2M_2)^{-1}=dE/dp^2$ arises from $\slsh{D}^2$ and the Wilson
term, whereas the chromomagnetic term arises only from $\slsh{D}^2$.
This leads to a mismatch with $(2M_3)^{-1}$, which is removed by the
clover term \cite{She85}.

\section{A TOY MODEL}\label{toy}
To derive eq.~(\ref{M-expansion}) one must set up the transfer matrix
formalism.
First one separates the four-com\-pon\-ent fields into two-component
fields, using the appendix.
%
The manipulations become quite cumbersome \cite{Lue77}, so
consider instead a toy model with the most important properties:
\begin{equation}
\begin{array}{l}
S={\displaystyle \sum_t}
  \phi^\dagger_t(\partial_0+m)\phi_t
+ \chi^\dagger_t(\partial_0+m)\chi_t \\[1.0em] \hspace{3,5em}
+ ip \phi^\dagger_t\chi^\dagger_t - ip \chi_t\phi_t .
\end{array}
\end{equation}
Then
$\cD\psi\cD\bar{\psi}=
\prod_t d\phi_t^\dagger d\phi_t d\chi_t^\dagger d\chi_t$ and
\begin{equation}
\begin{array}{l}\label{factorize}
e^{-S}={\displaystyle \prod_t}
e^{-(1+m)(\phi^\dagger_t\phi_t+\chi^\dagger_t\chi_t)} \times \\[1.0em]
\hspace{5.0em} {\displaystyle \prod_t}
T(\phi^\dagger_{t+1},\chi^\dagger_{t+1};\phi_t,\chi_t),
\end{array}
\end{equation}
where
\begin{equation}
T(\phi^\dagger,\chi^\dagger;\phi,\chi)= e^{-ip \phi^\dagger\chi^\dagger}
e^{\phi^\dagger\phi+\chi^\dagger\chi} e^{ ip \chi\phi}.
\end{equation}

The rules of Grassman integration imply
$\{\hat{a},\hat{a}^\dagger\}=1=\{\hat{b},\hat{b}^\dagger\}$, where
\begin{equation}\label{normalization}
\begin{array}{l}
\hat{a}=\sqrt{1+m}\,\hat{\phi} \\[1.0em]
\hat{b}=\sqrt{1+m}\,\hat{\chi}
\end{array}
\end{equation}
are operators in Hilbert space \cite{Lue77}.
The Fock vacuum satisfies $\hat{a}|0\rangle=\hat{b}|0\rangle=0$.
The other states are $|q\rangle=\hat{a}^\dagger|0\rangle$,
$|\bar{q}\rangle=\hat{b}^\dagger|0\rangle$, and
$|q\bar{q}\rangle=\hat{a}^\dagger\hat{b}^\dagger|0\rangle$.
Up to Jacobian factors, the matrix elements of the transfer matrix are
the coefficients of monomials in (Grassman numbers) $a$, $a^\dagger$,
$b$, and $b^\dagger$, when $T(\phi^\dagger,\chi^\dagger;\phi,\chi)$ is
expressed as a polynomial.
Thus
$\langle q|\hat{T}|q\rangle=\langle\bar{q}|\hat{T}|\bar{q}\rangle=
1+m$, and in the neutral sector
\begin{equation}\label{transfer-matrix}
\langle i|\hat{T}|j\rangle=\left(
\begin{array}{cc}
(1+m)^2 & ip(1+m) \\
-ip(1+m) & 1+p^2
\end{array}
\right)
\end{equation}
and all other elements are zero.
This model can be solved exactly, by diagonalizing
$\langle i|\hat{T}|j\rangle$.
Upon reconstructing the operator form and expanding in $p$, one obtains
the analog of eq.~(\ref{M-expansion}).

\section{THE NORMALIZATION}\label{the-normalization}
$\hat{a}^\dagger$ and $\hat{b}^\dagger$ create normalized states,
$\hat{\phi}^\dagger$ and $\hat{\chi}^\dagger$ do not.
Applying eq.~(\ref{normalization}) to Wilson fermions,
with $m\rightarrow am_0-\half a^2\triangle^{(3)}$, one sees that
\begin{equation}
a(\vek{x})=
\left(1 + a m_0-\half a^2\triangle^{(3)}\right)^{1/2}\psi(\vek{x})
\end{equation}
is canonically normalized; in momentum space,
\begin{equation}
a(\vek{p})=
\left(1 + a m_0+\half a^2\hat{\vek{p}}^2\right)^{1/2}\psi(\vek{p}).
\end{equation}
At $\vek{p}=\veg{0}$ this factor is $e^{M_1/2}$.

\begin{figure}[b]
\epsfxsize=0.45\textwidth
\epsfbox{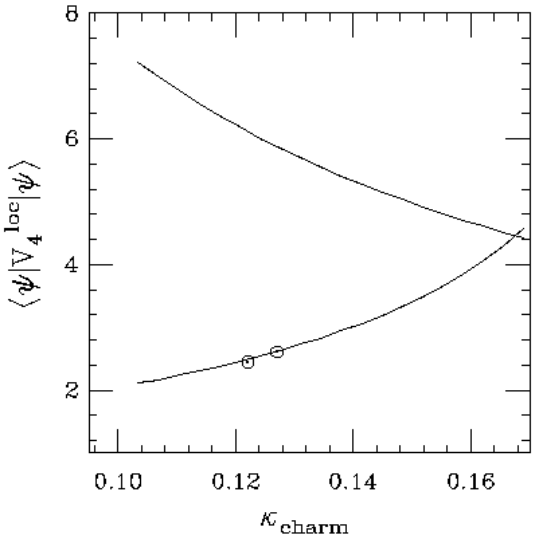}
\caption[Kha92]{
$\langle J/\psi|V_4|J/\psi\rangle$, where $V_4=\bar{\psi}\gamma_0\psi$
is the (unnormalized) local charge.
The lower line is the mean-field Ansatz
$\tilde{Z}_V/(1-6\tilde{\kappa})$
and the upper line is $Z_V/(2\kappa)$.
(The latter form is the one suggested by the massless limit.)
To get the charge (should be 1) divide the data by the curve.
}\label{Kha92}
\end{figure}

This fact has been noticed in several phenomenological analyses
\cite{Gup91,Ber92}, which suggested using
$e^{M/2}\bar{Q}(x)\gamma_\mu q(x)$ and similar bilinears to determine
heavy-light decay constants, using Ans\"atze such as
$e^M=1+1/(2\kappa)-1/(2\kappa_c)$.
Combining the transfer-matrix analysis with mean field theory suggests a
different Ansatz.
Including gauge fields and using the notation of
eq.~(\ref{Wilson-action})
\begin{equation}
a_n= (1 - \kappa B)^{1/2}_{nm} \psi_m,
\end{equation}
has unit normalization \cite{Lue77}, where
$B_{nm}\psi_m=\sum_i U_{n,i}\psi_{n+\hat{\imath}}+
U_{n-\hat{\imath},i}^\dagger\psi_{n-\hat{\imath}}$.
In mean-field theory one replaces the dynamical $U$'s by their average.
Hence $(1-6\tilde{\kappa})^{1/2}\psi_n$,
where $\tilde{\kappa}=\kappa\langle u\rangle=\kappa/8\kappa_c$
\cite{Lep92}, is expected to be correctly normalized to high accuracy.
As shown in fig.~\ref{Kha92} the mean-field Ansatz agrees perfectly
with numerical data for the local current.

\section{MORE ON THE NEW ACTION}\label{asymmetric-section}
This section concludes with two short comments on
eq.~(\ref{kappa-kappa}).
It is convenient to define
$am_0=(1-2\kappa_0-6\kappa_S)/(2\kappa_0)$ and
$\zeta=\kappa_S/\kappa_0$.
Then eq.~(\ref{M1}) and the denominator of eq.~(\ref{M2}) are the
same, but the numerator of eq.~(\ref{M2}) becomes
\begin{equation}
\zeta(2\zeta +2m_0a(1+\zeta) +m_0^2a^2).
\end{equation}
An appropriate choice of $\zeta$ imposes $M_1=M_2$.

The asymmetric action is also natural from the point of view of
renormalization.
Imagine starting with the Wilson action on a fine lattice, and imagine
performing a ``block spin'' to a coarse lattice.
In perturbation theory this entails calculating the self-energy
$\Sigma(p)$.
For a massive quark, the most natural subtraction point is on shell,
e.g.\ $E=M_1$, $\vek{p}=\veg{0}$.
On the lattice, however,
\begin{equation}
\left. \frac{\partial\Sigma}{\partial p_0} \right|_{\rm on~shell}
\neq
\left. \frac{\partial\Sigma}{\partial p_i} \right|_{\rm on~shell}
\end{equation}
In the hopping-parameter notation this difference implies
$\kappa_0\neq\kappa_S$ for the effective action on the coarse lattice.

\appendix

\section{FOUR TO TWO COMPONENTS}
The Dirac matrices are taken to be
\begin{equation}
\gamma_0=\left(
\begin{array}{cc}
1 & 0 \\ 0 & -1
\end{array}
\right),\hspace{2.0em}
\gamma_i=\eta\left(
\begin{array}{cc}
0 & \sigma_i \\ \sigma_i & 0
\end{array}
\right).
\end{equation}
In Minkowski space $\eta=-i$;
in Euclidean space $\eta=1$.
Fermion fields are decomposed into two-com\-pon\-ent form as follows:
\begin{equation}
\psi=\left(
\begin{array}{c}
\phi \\ \chi^\dagger
\end{array}
\right),\hspace{2.0em}
\bar{\psi}=(\phi^\dagger\;\;\;\;-\chi).
\end{equation}


\section*{ACKNOWLEDGEMENTS}
I would like to thank A.X. El-Khadra and P.B. Mackenzie for numerous
discussions on the material discussed in this talk.
Fermilab is operated by Universities Research Association, Inc.\ under
contract with the U.S. Department of Energy.


\begin{thebibliography}{99}
\bibitem{Wil77}
K.G. Wilson, in {\em New Phenomena in Subnuclear Physics}, edited by
A. Zichichi (Plenum, New York, 1977).
\bibitem{Eic87}
E. Eichten, Nucl. Phys. B Proc. Suppl. {\bf 4} (1987) 170;
E. Eichten and F. Feinberg, Phys. Rev. Lett. {\bf 43} (1979) 1205;
Phys. Rev. {\bf D23} (1981) 2724.
\bibitem{Lep87}
G.P. Lepage and B.A. Thacker,
Nucl. Phys. B Proc. Suppl. {\bf 4} (1987) 199;
B.A. Thacker and G.P. Lepage, Phys. Rev. {\bf D43} (1991) 196.
\bibitem{Kro92}
A.S. Kronfeld and P.B. Mackenzie, in progress.
\bibitem{She85}
B. Sheikholeslami and R. Wohlert, Nucl. Phys. {\bf B259} (1985) 572.
\bibitem{Hor92}
G.P Lepage, L. Magnea, C. Nakhleh, U. Magnea, and K. Hornbostel,
CLNS-92-1136.
\bibitem{Lue77}
M. L\"uscher, Commun. Math. Phys. {\bf 54} (1977) 283.
\bibitem{Gup91}
R. Gupta, C.F. Baillie, R.G. Brickner, G.W. Kilcup, A. Patel, and
S.R. Sharpe, Phys. Rev. {\bf D44} (1991) 3272.
\bibitem{Ber92}
C. Bernard, C.M. Heard, J. Labrenz, and A. Soni,
Nucl. Phys. B Proc. Suppl. {\bf 26} (1992) 384; this volume.
\bibitem{Lep92}
G.P. Lepage and P.B. Mackenzie, FERMILAB-PUB-91-355/T.
\end{thebibliography}
\end{document}